# Nitrogen Atmospheres of the Icy Bodies in the Solar System

**M. Scherf[1], H. Lammer[1], N.V. Erkaev[2,3], K.E. Mandt[4], S.E. Thaller[5], B. Marty[6]**

[1]Austrian Academy of Sciences, Space Research Institute, Schmiedlstr. 6, 8042 Graz, Austria (manuel.scherf@oeaw.ac.at, helmut.lammer@oeaw.ac.at);

[2]Institute of Computational Modeling, SB-RAS, 660036 Krasnoyarsk, Russian Federation (erkaev@icm.krasn.ru);

[3]Siberian Federal University, 660036 Krasnoyarsk, Russian Federation;

[4]Johns Hopkins University Applied Physics Laboratory, Laurel MD, USA;

[5]Institute of Physics/IGAM, University of Graz, Universitätsplatz 5, 8010 Graz, Austria;

[6]Centre de Recherches Pétrographiques et Géochimiques, UMR CNRS & Université de Lorraine, Vandoeuvre-les-Nancy Cedex, France.

**Abstract**   This brief review will discuss the current knowledge on the origin and evolution of the nitrogen atmospheres of the icy bodies in the solar system, particularly of Titan, Triton and Pluto. An important tool to analyse and understand the origin and evolution of these atmospheres can be found in the different isotopic signatures of their atmospheric constituents. The $^{14}N/^{15}N$ ratio of the $N_2$-dominated atmospheres of these bodies serve as a footprint of the building blocks from which Titan, Triton and Pluto originated and of the diverse fractionation processes that shaped these atmospheres over their entire evolution. Together with other measured isotopic and elemental ratios such as $^{12}C/^{13}C$ or Ar/N these atmospheres can give important insights into the history of the icy bodies in the solar system, the diverse processes that affect their $N_2$-dominated atmospheres, and the therewith connected solar activity evolution. Titan's gaseous envelope most likely originated from ammonia ices with possible contributions from refractory organics. Its isotopic signatures can yet be seen in the – compared to Earth – comparatively heavy $^{14}N/^{15}N$ ratio of 167.7, even though this value slightly evolved over its history due to atmospheric escape and photodissociation of $N_2$. The origin and evolution of Pluto's and Triton's tenuous nitrogen atmospheres remain unclear, even though it might be likely that their atmospheres originated from the protosolar nebula or from comets. An in-situ space mission to Triton such as the recently proposed Trident mission, and/or to the ice giants would be a crucial cornerstone for a better understanding of the origin and evolution of the icy bodies in the outer solar system and their atmospheres in general. Due to the importance of the isotopic measurements for understanding the origin and evolution of the icy bodies in the solar system, this review will also give a brief discussion on the diverse measurement techniques with a focus on nitrogen.

# 1 Introduction

Only a small fraction of the more than 190 satellites in the solar system are currently known to have tenuous atmospheres (see e.g., Coustenis et al., 2010). These atmospheres are dominated by different species such as oxygen at Europa (Hall et al., 1995, 1998) or Ganymede (Noll et al., 1996; Hall et al., 1998; Feldman et al., 2000), $CO_2$ (Carlson, 1999), CO, $O_2$, O and C in Callisto's exosphere (Vorburger et al., 2019), water vapor at Enceladus (Dougherty et al., 2006), $SO_2$ at Io (Pearl et al., 1979; Lellouch et al., 2007) or tenuous nitrogen atmospheres at Triton and Pluto (Smith et al. 1989; Gladstone et al. 2016; Gladstone and Young 2019; Hart 1974). Among all known satellites in the solar system only Titan has a thick atmosphere with nitrogen as its main constituent. This makes Titan, Triton and Pluto the only icy bodies in the solar system with an $N_2$-dominated atmosphere.

Besides these icy bodies, only Earth's atmosphere is in addition dominated by nitrogen which likely originated from chondritic meteorites (Marty, 2012; Harries et al., 2015). The origin – but also the evolution – of $N_2$ within the atmospheres of the icy bodies might be a different one. An indication for this can be found in the isotopic signature of $N_2$ which gives clues on the origin and evolution of the respective atmospheres. While Earth's atmospheric nitrogen has an isotopic ratio of $^{14}N/^{15}N \sim 272$ (Marty and Zimmermann, 1999; see also Table 1; Füri et al., 2015), which is close to chondritic, Titan's value is $167.7 \pm 0.3$ (Niemann et al., 2010); for Pluto and Triton this is currently unknown. The difference between the terrestrial and Titan's $^{14}N/^{15}N$ ratio can be explained either by different building blocks or by fractionation processes that changed the respective ratios over time – or by both.

Table 1. $^{14}N/^{15}N$ throughout the solar system.

| Solar system object | $^{14}N/^{15}N$ | Reference |
|---|---|---|
| Sun/Protosolar Nebula | 440.5±5.8 | Marty et al. 2011 |
| Solar wind | 458.7±4.1 | Marty et al. 2011 |
| Venus' atmosphere | 273±56 | Hoffman et al., 1979 |
| Earth's atmosphere | 272±0.3 | Marty and Zimmermann 1999 |
| Earth's mantle | 273.4±0.8 | Cartigny, 2005 |
| Earth's primordial mantle | ~283 | Cartigny and Marty, 2013 |
| Mars' atmosphere | 168±17 | Wong et al. 2013 |
| Mars' interior | ~280 | Mathew and Marti, 2001 |
| Jupiter | 435±65 | Owen et al. 2001 |
| Saturn | >357 | Fletcher et al. 2014 |
| Titan (present-day) | 167.7±0.6 | Niemann et al. 2010 |
| Titan (past) | ≤190; | Mandt et al. 2014; |
|  | ≤129; | Krasnopolsky 2016; |
|  | ≤166 – 172 | Erkaev et al. 2020 |
| Comet C/2012 S1 (ISON) | 139±38 | Shinnaka et al. 2014 |
| NH$_2$ of Jupiter family and Oort cloud comets | 127±32 | Rousselot et al. 2014 |
| NH$_2$ of additional 18 comets from various dynamical groups | 135.7±5.9 | Shinnaka et al. 2016 |
| Refractory organics | 160-320 (mean: 231) | Miller et al., 2019 |
| Chondrites | 259±15 | Alexander et al. 2012 |
| Triton | unknown | |
| Pluto | unknown | |

There are several primordial nitrogen reservoirs in the solar system with distinct $^{14}$N/$^{15}$N ratios from which the building blocks of Titan's – and of Triton's and Pluto's – N$_2$-atmosphere might have originated (see also Table 1). These reservoirs include

i) chondrites (e.g., Kerridge, 1985; Alexander et al., 2012), the likely reservoir of the terrestrial nitrogen (Marty and Zimmermann, 1999; Marty, 2012),

ii) the solar wind and Jupiter, which are believed to represent N$_2$ within the protosolar nebula (e.g., Meibom et al., 2007; Marty et al., 2010),

iii) ammonia ices (Rousselot et al., 2014; Shinnaka et al., 2014a, 2016a; Shinnaka and Kawakita, 2016), cyanide (CN) and hydrogen cyanide (HCN; e.g., Bockelée-Morvan et al., 2008; Manfroid et al., 2009) in comets, and

iv) complex refractory organics which are observed in cometary grains (e.g., K. D. McKeegan et al., 2006; De Gregorio et al., 2010), interplanetary dust particles (IDPs; e.g., Messenger, 2000; Aléon et al., 2003; Kevin D. McKeegan et al., 2006; Busemann et al., 2009), and as insoluble organic matter (IOM) in primitive chondrites (e.g., Aléon, 2010; Alexander et al., 2017).

Since organic matter, however, is the main carrier of nitrogen within chondrites (e.g., Aléon, 2010), these may not be classified as an independent reservoir but as an admixture of organics with another nitrogen-bearing reservoir.

By reconstructing the initial fractionation of Titan's building blocks one can determine the specific reservoir(s) out of which Titan's $N_2$ originated. Various processes, however, can fractionate isotopic ratios and must be considered to determine the initial $^{14}N/^{15}N$ ratio of a satellite or a planet. While in the specific case of Earth biological, geological, and anthropogenic processes are also fractionating nitrogen, it is expected that the main drivers for fractionation on the icy bodies are

i)   atmospheric escape (Lammer et al., 2002; Lammer and Bauer, 2003; Kathleen E. Mandt et al., 2009; Mandt et al., 2014) and

ii)  photochemical processes such as photodissociation of $N_2$ (Liang et al., 2007; Krasnopolsky, 2016), even though this process might be more important on Titan than on Pluto (Mandt et al., 2017).

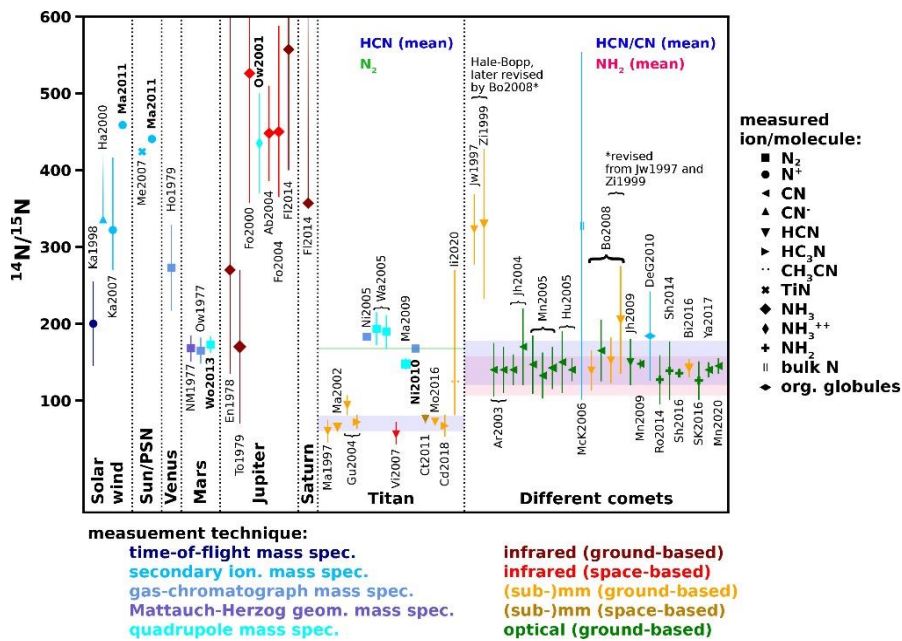

**Fig. 1. Measurements of $^{14}N/^{15}N$ throughout the solar system (except for the Earth, interplanetary dust particles, and chondrites which are generally measured with high precision in sophisticated Earth-based laboratories) for different molecules and measurement techniques including sector-type mass spectrometry and secondary ionization**

mass spectrometry for sample return missions. The bluish and reddish areas illustrate the mean values for HCN(/CN) and $NH_2$ for Titan and comets, respectively, while the greenish area just illustrates the value of $N_2$ at Titan as a reference. The bold references are the current standards for these respective ratios. References are Ka1998 (Kallenbach et al., 2003), Ha2000 (Hashizume et al., 2000), Ka2007 (Kallenbach et al., 2007; re-evalutation of Ka1998), Ma2011 (Marty et al., 2011), Me2007 (Meibom et al., 2007), Ho1979 (Hoffman et al., 1979), NM1977 (Nier and McElroy, 1977), Ow1977 (Owen et al., 1977), Wo2013 (Wong et al., 2013), En1978 (Encrenaz et al., 1978), To1979 (Tokunaga et al., 1979), Fo2000 (Fouchet et al., 2000), Ow2001 (Owen et al., 2001b), Ab2004 (Abbas et al., 2004), Fo2004 (Fouchet et al., 2004), Fl2014 (Fletcher et al., 2014), Ma1997 (Marten et al., 1997), Ma2002 (Marten et al., 2002), Gu2004 (Gurwell, 2004), Ni2005 (Niemann et al., 2005), Wa2005 (Waite et al., 2005), Ma2009 (Kathleen E Mandt et al., 2009), Ni2010 (Niemann et al., 2010), Ct2010 (Courtin et al., 2011), Mo2016 (Molter et al., 2016), Cd2018 (Cordiner et al., 2018), Ii2020 (Iino et al., 2020), Jw1997 (Jewitt et al., 1997), Zi1999 (Ziurys et al., 1999), Ar2003 (Arpigny et al., 2003), Jh2004 (Jehin et al., 2004), Mn2005 (Manfroid et al., 2005), Hu2005 (Hutsemékers et al., 2005), McK2006 (McKeegan et al., 2006), Bo2008 (Bockelée-Morvan et al., 2008), Jh2009 (Jehin et al., 2009), Mn2009 (Manfroid et al., 2009), DeG2010 (De Gregorio et al., 2010), Ro2014 (Rousselot et al., 2014), Sh2014 (Shinnaka et al., 2014b), Sh2016 (Shinnaka et al., 2016b), Bi2016 (Biver et al., 2016), SK2016 (Shinnaka and Kawakita, 2016), Ya2017 (Yang et al., 2018), Mn2020(Moulane et al., 2020)

While atmospheric escape preferentially removes the lighter isotope from the atmosphere (see also Lammer et al. (2020), this issue, and Lammer et al. (2008) for a review on fractionation by atmospheric escape), photolysis of $N_2$ preferentially removes the heavier isotope (Liang et al., 2007). All these processes must be taken into account to gain insights into the origin and evolution of the nitrogen dominated atmospheres of Titan, Triton and Pluto. This brief review will mainly focus on these aspects while discussing the current knowledge of the nitrogen atmospheres around these icy bodies. In Section 2, we will further review different measurement techniques for obtaining the isotopic composition of nitrogen throughout the solar system, since its measurement accuracy – and our therewith derived scientific knowledge from these icy bodies – strongly relies on the involved technology. Section 3 will discuss Titan's nitrogen atmosphere and Section 4 will give an overview on the tenuous atmospheres of Pluto and Triton. A discussion and conclusion finalize the review.

## 2 Nitrogen Isotopes Measurement Techniques

The isotopic ratio of $^{14}N/^{15}N$ on Titan was long believed to be around $60 - 70$ as determined through remote measurements of HCN in its atmosphere (Marten et al., 1997, 2002; Gurwell, 2004). However, as the space mission Cassini-Huygens has subsequently illustrated, the isotopic ratio of Titan's $N_2$ cannot be derived adequately by just measuring nitrogen-bearing molecules such as HCN. This further exemplifies the difficulties in determining isotopic ratios of $N_2$ throughout the Solar System. It is, therefore, crucial to comprehend how these isotopic ratios were obtained, what is its measurement methods and the difficulties connected with these techniques. These will be addressed in this chapter.

Determining the abundance and ratio of the stable isotopes of nitrogen, $^{14}N$ and $^{15}N$, throughout the solar system can principally be performed either through mass spectrometry or high-resolution spectroscopy ranging from the optical up to sub-millimetre wavelengths. Measurements of these ratios for different N-bearing molecules, including the involved measurement techniques for different solar system bodies, are illustrated in Fig. 1.

In case of mass spectrometry, the instrument has either to be transported to the target or the target to a laboratory on Earth. Besides a few sample return missions (e.g., Kallenbach et al., 1998; McKeegan et al., 2006; Marty et al., 2011) or investigations of primitive meteorites (e.g., Kerridge, 1985; Busemann et al., 2006; Alexander et al., 2007; Alexander, 2011) and IDPs (e.g., Messenger, 2000; Floss et al., 2006; Busemann et al., 2009) that, among other research questions, also investigated the nitrogen isotopic composition of its target, mass spectrometers will have to be adopted to be suitable for transportation to the sample, such as the nitrogen-dominated atmosphere of Titan. The respective instrument has to be miniaturized from its complex laboratory setup to the small-sized, low-weight requirements of a planetary mission, thereby inevitably reducing its sensitivity and resolution (for reviews on mass spectrometry in planetary sciences see e.g., Palmer and Limero, 2001; Ireland, 2013; Ren et al., 2018; Yokota, 2018; McIntyre et al., 2019; Arevalo et al., 2020).

Consequently, mass interferences between different isotopologues of similar mass significantly complicate the determination of isotopic ratios and increase its error bars. The Galileo Probe Mass Spectrometer (GPMS; Niemann et al., 1992), a quadrupole mass spectrometer (see e.g., Arevalo et al., 2020, for a thorough description of different types of mass spectrometers adopted for space missions), measured the Jovian $^{14}$N/$^{15}$N ratio from the doubly charged $^{14}$NH$_3^{++}$ and $^{15}$NH$_3^{++}$ signals (see also Fig. 1) at 9 and 8.5 m/z, since H$_2$$^{16}$O at 18 m/z interfered with $^{15}$NH$_3^+$, making a measurement through $^{14}$NH$_3^+$ and $^{15}$NH$_3^+$ not possible (Wong et al., 2004). Similarly, the Mars Science Laboratory SAM/Quadrupole Mass Spectrometer (QMS; Mahaffy et al., 2012) could not infer the Martian $^{14}$N/$^{15}$N through direct atmospheric experiments from N$_2^+$ due to interferences of CO$^+$ ions from CO$_2$ but only from N$^+$ ions or through enrichment experiments where CO$_2$ and H$_2$O were removed by chemical scrubbers (Wong et al., 2013).

However, in-situ mass spectrometry, besides sample return, nevertheless provides the most accurate means for direct measuring of nitrogen isotopes on distant solar system bodies, if the instrument's resolution allows to bypass interferences, and its sensitivity for a detection of the less abundant isotopologue. Another inaccuracy introduced by mass spectrometry might be coming from the spatial area of the measurement, i.e. where the measurements were performed. The Ion and Neutral Mass Spectrometer (INMS) of Cassini (Waite et al., 2004) determined the $^{14}$N/$^{15}$N ratio at Titan by crossings through its tenuous upper atmosphere above the homopause. Due to diffusive separation between the lighter and heavier isotopes above the well-mixed lower atmosphere, the measured values (Waite et al., 2005) from Cassini/INMS of $^{14}$N/$^{15}$N = $188^{+14}_{-16}$, as determined through the integrated sums of the measured abundances below 1230 km, and $^{14}$N/$^{15}$N = $214 \pm 1$, as retrieved through mass deconvolution of the average spectra between the height of 1230 and 1180 km, must be extrapolated to the surface, thereby necessarily introducing further inaccuracies. Waite et al. (2005) extrapolated these values to be $^{14}$N/$^{15}$N = 168 – 211 on Titan's surface, while Mandt et al. (2009) estimated the surface value based on Cassini/INMS data to be $^{14}$N/$^{15}$N = $148 \pm 7.5$.

Another instrument on-board Cassini-Huygens, the Gas-Chromatograph Mass Spectrometer (GCMS; Niemann et al., 2002) measured the value directly at the surface to preliminarily be $^{14}N/^{15}N = 183 \pm 5$ directly from molecular nitrogen (Niemann et al., 2005). This was later re-evaluated to $167.7 \pm 0.6$ due to a refined instrument calibration (Niemann et al., 2010) and is now considered as the standard-value for Titan's atmosphere.

**Table 2.** Measurements and techniques of $^{14}N/^{15}N$ at Titan.

| species | $^{14}N/^{15}N$ ratio | reference | technique | instrument | remarks |
|---|---|---|---|---|---|
| HCN | $60 \pm 15$ | Marten et al. 1997 | (sub-)mm (ground) | IRAM 30-m telescope (Baars et al. 1987) | |
| HCN | $65 \pm 5$ | Marten et al. 2002 | (sub-)mm (ground) | IRAM 30-m telescope (Baars et al. 1987) | renewal of telescope & receiver in 1998; better position of Saturn |
| HCN, HC₃N | $94 \pm 13$ (a) $72 \pm 9$ (b) | Gurwell 2004 | (sub-)mm (ground) | Submillimetre Array | temperature profiles of (a) Coustenis & Bezard 1995, and (b) Lellouch 1990 |
| N₂ | $183 \pm 5$ | Niemann et al. 2005 | gcms[a] | Huygens/GCSM (Niemann et al. 2002) | preliminary results |
| N₂ | 172-215 (a) 168-211 (b) | Waite et al. 2005 | qms[b] | Cassini/INMS (Waite et al. 2004) | (a) measured above homopause (b) extrapolated to surface |
| HCN | $56^{+16}_{-13}$ | Vinatier et al. 2007 | IR (space) | Cassini/CIRS (Flasar et al. 2004) | |
| N₂ | $147.5 \pm 7.5$ | Mandt et al. 2009 | qms[b] | Cassini/INMS (Waite et al. 2004) | extrapolated to surface |
| N₂ | $167.7 \pm 0.6$ | Niemann et al. 2010 | gcms[a] | Huygens/GCSM (Niemann et al. 2002) | refined results based on better laboratory calibration |
| HCN | $76 \pm 6$ | Courtin et al. 2011 | (sub-)mm (space) | Herschel-SPIRE (Griffin et al. 2010) | |
| HCN | $72.3 \pm 2.2$ | Molter et al. 2016 | (sub-)mm (ground) | ALMA (see Molter et al. 2016) | |
| HC₃N | $67 \pm 14$ | Cordiner et al. 2018 | (sub-)mm (ground) | ALMA (see Cordiner et al. 2018) | |
| CH₃CN | $125^{+145}_{-44}$ | Iino et al. 2020 | (sub-)mm (ground) | ALMA (see Iino et al. 2020) | |

[a]gas-chromatograph spectrometer; [b]quadrupole mass spectrometer

Another possibility for in-situ determination of isotopic ratios is the highly-sensitive Tunable Laser Spectroscopy (TLS) which is currently part of the SAM instrument-suite (Mahaffy et al., 2012) on-board of the Mars Science Laboratory (MSL). TLS provides sensitivities of ≈1‰ for planetary low-mass instruments (e.g., Webster et al., 2014) and works through spectroscopy in the IR region with a tunable diode laser light source that can be tuned over the specific molecular absorption line wavelengths of interest, thereby being absorbed by the respective molecule. While this method is well suited for the characterization of $H_2O$, CO, $CO_2$, $CH_4$, or $O_3$, and indeed recorded high-precision stable isotope ratios on Mars for hydrogen, oxygen and carbon isotopes (e.g., Webster et al., 2013), it is less appropriate for the characterization of $N_2$. Molecular nitrogen does not show significant absorption lines in the IR region (e.g., Lofthus and Krupenie, 1977). However, TLS is principally also suited to observe NO, $NO_2$, $HNO_3$, and $NH_3$ (Webster et al., 2014) and was already proposed for a mission to Titan (Webster et al., 1990), and, more recently, to Venus (Webster et al., 2015).

That molecular nitrogen only shows very weak absorption lines is also a problem for spectroscopic observations from ground and space, since spectroscopic techniques that are normally used to measure other isotopic ratios cannot be applied to $N_2$. This also prohibits a remote determination of the $^{14}N/^{15}N$ ratio in Titan's atmosphere through spectroscopic observations of $N_2$. However, other nitrogen-bearing molecules show spectral features and can be observed from Earth and from space, which has been performed in the past for several different molecular species in Titan's atmosphere, as can be seen in Table 2.

Depending on the molecule, fingerprints of nitrogen can be observed in the optical, such as the molecules $NH_2$, a primer for $NH_3$ (Rousselot et al., 2014; Shinnaka et al., 2014a, 2016a; Shinnaka and Kawakita, 2016) and CN (Arpigny et al., 2003; Jehin et al., 2004, 2009; Hutsemékers et al., 2005; Manfroid et al., 2005, 2009; Bockelée-Morvan et al., 2008; Yang et al., 2018; Moulane et al., 2020) as both observed for different comets, HCN (Vinatier et al., 2007, for Titan) and $NH_3$ (Encrenaz et al., 1978; Tokunaga et al., 1979; Fouchet et al., 2000, 2004; Abbas et al., 2004; Fletcher et al., 2014, all for Jupiter) in the infrared, and HCN, $HC_3N$, and $CHC_3N$ for Titan (Marten et al., 1997, 2002; Gurwell, 2004; Courtin et al., 2011; Molter et al., 2016; Cordiner et al., 2018; Iino et al., 2020) and for comets (Jewitt et al., 1997; Ziurys et al., 1999; Bockelée-Morvan et al., 2008; Biver et al., 2016) in the sub- to millimetre wavelength range.

A few space-bourne measurements of some of these molecules have been performed from afar and through flybys. During the latter, Cassini's Composite Infrared Spectrometer (CIRS; Flasar et al., 2004) performed measurements on HCN at Titan (Vinatier et al., 2007) and $NH_3$ at Jupiter (Abbas et al., 2004; Fouchet et al., 2004). These were the only successful attempts so far to determine nitrogen isotopic ratios in-situ with spacecraft instrumentation. For Saturn, Cassini/CIRS was unable to determine $^{14}N/^{15}N$ through $NH_3$. Due to the giant planet's lower atmospheric temperature and the dominance of phosphine in its atmosphere in the mid-infrared, the sensitivity requirements were too high for Cassini/CIRS (e.g., Fletcher et al., 2014).

Other space-based spectroscopic determinations of $^{14}N/^{15}N$ were performed from the distance; infrared observations of $NH_3$ at Jupiter by Fouchet et al. (2000) with ISO-SWS (De Graauw et al., 1996) and submillimetre observations of HCN with Herschel-SPIRE (Griffin et al., 2010) by Courtin et al. (2011). Several different measurements of $^{14}N/^{15}N$ ratios on a diverse set of solar system bodies were additionally performed from the ground. While ground-based instrumentation has the advantage of not having the need to be miniaturized as for a space-based detector, interferences and low transmission rates due to e.g. $H_2O$ and $O_3$ in the Earth's atmosphere, as well as local weather conditions, complicate observations from the ground. However, the first determination of the $^{14}N/^{15}N$ ratio in Titan's atmosphere was performed with the IRAM 30-m telescope (Baars et al., 1987) by Marten et al. (1997), who measured the ratio in HCN to be $^{14}N/^{15}N = 65 \pm 15$. This was later refined by Marten et al. (2002) to $^{14}N/^{15}N = 65 \pm 5$, which was made possible due to higher sensitivity observations with upgraded receivers and a renewed IRAM 30-m telescope, but also due to a better position of Saturn in the sky as for their previous observations.

At the time of these measurements, the nitrogen isotopic ratio of HCN was expected to represent Titan's atmospheric $N_2$ as well. While other subsequent observations of HCN by Gurwell (2004) more or less confirmed the results of Marten et al. (2002), preliminary measurements of Cassini-Huygens/GCSM found an $^{14}N/^{15}N$ ratio for $N_2$ that was much higher, i.e. $^{14}N/^{15}N = 183 \pm 5$ (Niemann et al., 2005). This discrepancy was later to be resolved by photolytic fractionation of $^{14}N^{14}N$ and $^{14}N^{15}N$ (Liang et al., 2007) from $N_2$, which leads to a preferential incorporation of $^{15}N$ into HCN, thereby explaining the different isotopic ratios of $N_2$ and HCN, one of its daughter products.

This, however, also illustrates the difficulty to determine the ratio of $^{14}N$ and $^{15}N$ in $N_2$ through measurements of other molecules such as HCN, CN, or $NH_2$, which all show lower $^{14}N/^{15}N$ ratios than molecular nitrogen. Further problems in determining isotopic ratios from spectral observations come from missing experimental determinations of the spectral lines of particularly the less abundant isotopes (e.g., Rousselot et al., 2014), but also from theoretical models that are needed to translate emission lines into abundances and isotopic ratios. Derived atmospheric temperature profiles, for instance, significantly influence emission lines and the subsequent determination of isotopic abundances. Gurwell (2004) retrieved two slightly diverging values for the $^{14}N/^{15}N$ ratio in Titan's HCN due to the usage of different temperature profiles. They determined $^{14}N/^{15}N$ to be $94 \pm 13$ for a profile provided by Coustenis and Bezard (1995), and $^{14}N/^{15}N = 72 \pm 9$ for another one by Lellouch (1990).

Furthermore, observations of the $^{14}N/^{15}N$ ratio in ammonia ices in comets only became possible due to experimental determination of the $^{15}NH_2$ emission lines by Rousselot et al. (2014), which serves as proxy for the very faint lines of $^{15}NH_3$. Likewise, the fractionation between HCN and $N_2$ could have only been determined due to quantum-mechanical advances in the computation of isotopic photoabsorption and photodissociation cross sections of $N_2$ (Liang et al., 2007). This fractionation process further turns out to be dependent on the incident EUV flux (Mandt et al., 2017).

These examples demonstrate the difficulties for determining accurate $^{14}N/^{15}N$ ratios for any N-bearing molecules but in particular for $N_2$ through the extrapolation from CN, HCN, or $NH_2$. It also provides good arguments for in-situ measurements with mass spectrometers or for sample return missions, such as the Stardust mission (McKeegan et al., 2006; De Gregorio et al., 2010), or Genesis (Marty et al., 2011). The latter returned a sample of solar wind $N^+$ ions to be analysed via a high-precision technique called secondary ionization mass spectrometry (SIMS; see e.g., Schaepe et al., 2019, for an introduction into this technique) at the Cameca 1280HR2 instrument at CRPG Nancy (Marty et al., 2011), and confirmed the highly $^{14}N$-enriched (proto-)solar reservoir out of which Jupiter accreted its nitrogen. The sample return of the Stardust mission from comet 81P/Wild 2 confirmed the existence of nitrogen-bearing insoluble organic matter in cometary dust (McKeegan et al. 2006). This was as well investigated with the SIMS technique, but by the nanoSIMS instrument of the Washington University in St. Louis. Another high precision measurement technique, the so-called sector-type mass spectrometry, which either uses a static electric or a magnetic sector, or a combination of both, to retrieve high resolution and reproducibility measurements, was applied on the ground for analyzing noble gases in the Genesis targets (e.g., Meshik et al., 2014, 2020), and for analyzing Martian atmospheric nitrogen and noble gases trapped in Martian meteorites with high precision, supplementing the in-situ measurements done by the Viking and the MSL mass spectrometers (e.g., Becker and Pepin, 1984; Cassata, 2017). This technique was further also applied for determining the evolution of atmospheric xenon and other noble gas isotopes over the Earth's Archean eon (see, Avice et al., 2018; Avice and Marty, 2020, this issue).

As this overview of different measurement techniques for the determination of the $^{14}N/^{15}N$ ratios of diverse solar system bodies illustrates, one must be cautious in extrapolating a specific measurement to another species or even solar system body. While there are already several determinations of $^{14}N/^{15}N$ existing for different molecular species at Titan, these results cannot simply be generalized to isotopic nitrogen ratios at Triton and Pluto. As the following chapters will show, determining their isotopic ratios independently in the future will be crucial for better understanding the origin and evolution of their nitrogen atmospheres.

## 3 Titan

Saturn's large satellite Titan (Fig. 2; Table 1 and 2) is the only moon in the solar system with a substantial gaseous envelope, and besides Earth the only body with a thick $N_2$-dominated atmosphere. It has a surface pressure of 1.5 bar and consists of 98.4% $N_2$ and 1.5% $CH_4$ in the lower stratosphere (Niemann et al., 2010); at the surface the methane fraction even rises up to 5.7% (Samuelson et al., 1997; Niemann et al., 2010). While the existence of its atmosphere was first suggested already in 1908 (Solá, 1908) and proven in 1944 via the discovery of methane (Kuiper, 1944), $N_2$ being its main constituent was not confirmed until the flybys of Voyager 1 and 2 in the early 1980s (Broadfoot et al. 1981; Lindal et al. 1983; Hunten et al., 1984; Vervack et al. 2004) due to the challenging detection and confirmation of $N_2$ (see Section 2).

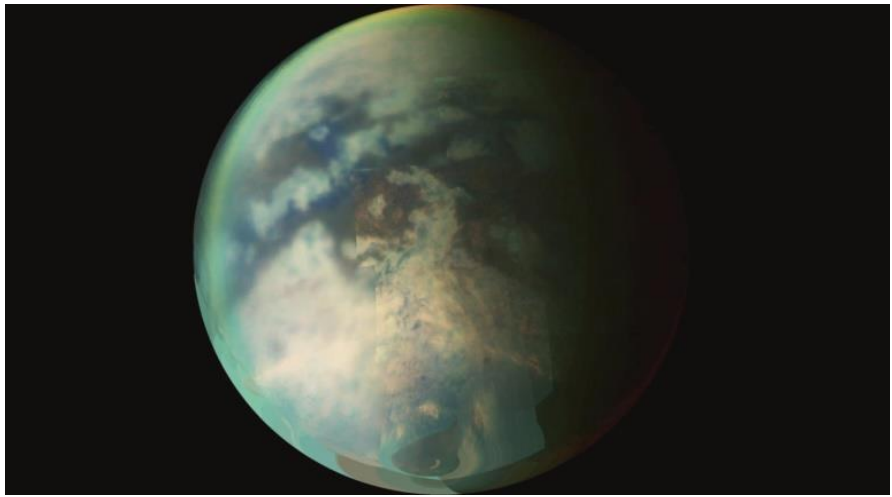

**Fig. 2. Cassini image of Saturn's moon Titan (Courtesy NASA/JPL/University of Arizona).**

The origin of nitrogen in Titan's atmosphere, however, is still a matter of debate (Niemann et al., 2010) with its most likely sources being proposed to be either $N_2$ or $NH_3$ (e.g. Strobel, 1982). Lewis (1971), who first suggested that Titan possesses a thick $N_2$-atmosphere, proposed that it accreted from $NH_3$, i.e. ammonia, during the formation of the moon. This ammonia was then either decomposed to $N_2$ via photolysis (Lewis, 1971; Atreya et al., 1978) or impact-induced chemistry (McKay et al., 1988; Ishimaru et al., 2011; Sekine et al., 2011), or outgassed as $N_2$ through thermal decomposition from Titan's interior (Glein et al., 2009; Glein, 2015). If $NH_3$ might have been efficiently altered to $N_2$ via impacts into an early ammonium hydrate rich crust (Sekine et al., 2011), such impacts would have also led to significant atmospheric erosion (Marounina et al., 2015).

As described in Section 2, first measurements of Titan's atmospheric $^{14}N/^{15}N$-ratio (see Table 2) suggested a strong enrichment of $^{15}N$ by about 4.5 times (Marten et al., 1997; Lunine et al., 1999) compared to the terrestrial atmospheric value of ~272 (Marty and Zimmermann, 1999). Since the conversion processes of $NH_3$ to $N_2$ could not have significantly altered the $^{14}N/^{15}N$-ratio (Berezhnoi, 2010), it was proposed that, in case Titan's atmosphere would have the same building blocks than the Earth's atmospheric nitrogen, this enrichment might indicate strong atmospheric loss of $N_2$, most likely through atmospheric escape early in Titan's history (e.g. Lunine et al., 1999; Lammer et al., 2000, 2008; Penz et al., 2005). These results also suggested that Titan's volatile inventory was accreted at least partially from the Saturnian subnebula which was more reducing than the surrounding solar nebula (Lunine et al., 1999). The $^{14}N/^{15}N$-ratio, however, was more precisely measured by the Cassini-Huygens/GCMS to be 167.7 (see Tables 1 and 2; Niemann et al., 2010) which was found to be close to the $^{14}N/^{15}N$-ratios of cometary ammonia (Rousselot et al., 2014; Shinnaka et al., 2014, 2016).

As has been shown by Mandt et al. (2009, 2014) atmospheric loss through hydrodynamic escape, sputtering and Jeans escape cannot explain the strong enrichment of $^{15}N$ in Titan's atmosphere compared to the terrestrial value, since all of these processes would have required a longer duration than the existence of the solar system for fractionating Titan's value from 272 to 167.7. Mandt et al. (2014) therefore concluded that the most likely source of Titan's $N_2$-atmosphere must have been ammonia ices with an initial ratio of <190 (see Fig. 3). This result was recently

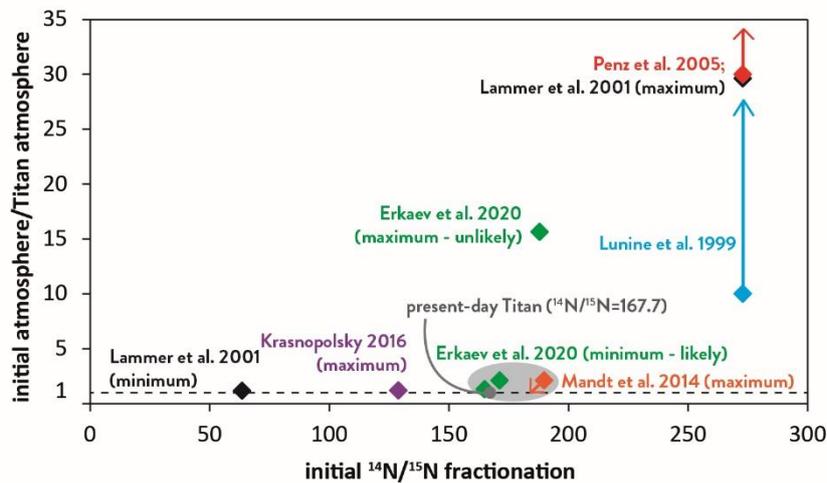

**Fig. 3. Different attempts to reproduce the initial $^{14}N/^{15}N$-fractionation of Titan. While older studies like Lunine et al (1999) or Penz et al. (2005) attempted to reproduce the terrestrial fractionation (i.e. ~273) and therefore yielded unrealistically high losses, newer studies calculated the loss through time to yield a maximum initial fractionation. The grey shaded area most likely represents the initial ratios.**

confirmed by a new study of Erkaev et al. (2020) who applied a hydrodynamic upper atmosphere model (Erkaev et al., 2015) that solves the equations of mass, momentum and energy for Titan's $N_2$-atmosphere under different EUV-fluxes from the young Sun. These authors found that depending on the solar EUV evolution, the $^{14}N/^{15}N$-ratio of Titan's atmosphere could have only been fractionated by thermal and non-thermal atmospheric escape and photochemical processes at maximum from an initial value of 167 for a weakly active (i.e. a so-called "slow rotator") to 172 for a moderately active, and 185 for a highly active (fast rotating) young Sun according to Tu et al. (2015). The total loss of nitrogen to space, if one assumes that Titan's atmosphere originated ~4.5 Gyr ago, varies significantly in their study. While for a slow and moderate rotator the total loss would only have been ~0.6 bar, and ~2 bar, respectively, it would have increased to

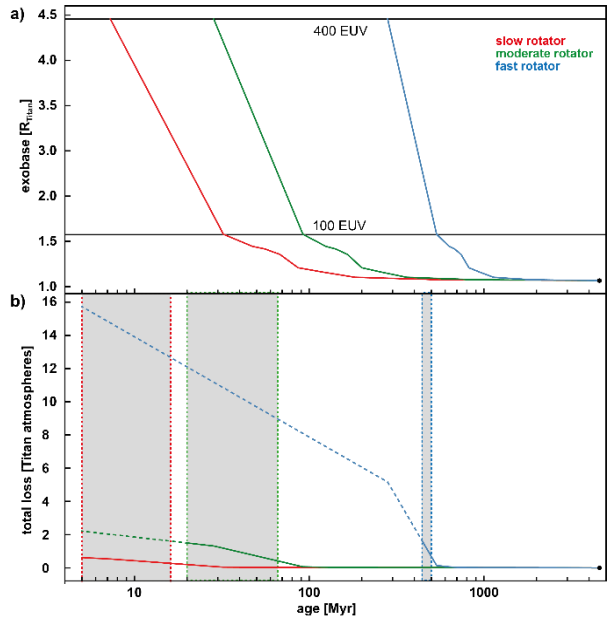

**Fig. 4. Upper panel: Exobase levels of Titan's nitrogen atmosphere for different EUV fluxes for a slow, moderate, and fast rotating Sun according to Erkaev et al. 2020. Lower panel: Total cumulative loss of Titan's atmosphere over time in units of Titan atmospheres. The grey areas surrounded by the respective dotted lines correspond to the earliest possible outgassing of Titan's atmosphere for a slow, moderate and fast rotator, respectively, based on the model of Glein (2015) but with the escape of Erkaev et al. (2020) included. (After Erkaev et al. 2020).**

~16 bar for a fast rotating Sun (Erkaev et al. 2020). For high EUV fluxes, a nitrogen-dominated atmosphere with negligible abundances of $CO_2$ will start to rapidly expand and hydrodynamically escape to space (see e.g., Erkaev et al. 2020; Johnstone et al. 2019; Tian et al. 2009). To experience such a behaviour, the irradiation onto Titan's atmosphere must have reached values of above 100 times the present-day EUV flux (Erkaev et al. 2020). Even though the young Sun most probably reached such high values very early in its history for a few million years, only a fast rotating Sun could have maintained these high fluxes for a significant period of time (Tu et al. 2015; Güdel et al. 2020; this issue).

Fig. 4a shows the exobase altitudes of Titan's nitrogen atmosphere for different solar EUV fluxes for a slow, moderate and fast rotator, respectively, as simulated with the model of Erkaev et al. (2020); Fig. 4b shows the related cumulative loss in Titan's nitrogen atmospheres over time. Contrary to the present-day, for which sputtering is the dominant loss process, thermal escape would have clearly dominated early in Titan's history (with escape rates up to $10^{30}$ s$^{-1}$ for 400 EUV), in case that its nitrogen-dominated atmosphere was already present at that time.

If, however, the proposed endogenic origin of Titan's atmosphere through decomposition of NH$_3$ and subsequent outgassing as N$_2$, as proposed by Glein (2015), is

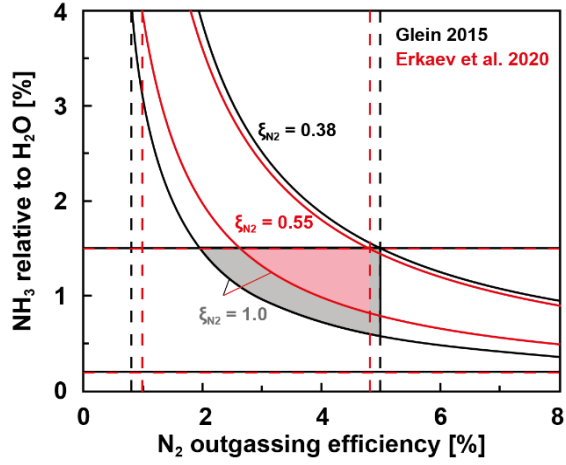

**Fig. 5. The parameter space for making Titan's nitrogen atmosphere according to Glein (2015) and, for a slow rotator, to Erkaev et al. (2020). Here it is assumed that the atmosphere originated endogenically from geochemical synthesis of NH$_3$ into N$_2$ within the deep interior of Titan; $\xi_{N2}$ = 1 corresponds to 100% synthesis. The dashed lines illustrated limiting values according to Glein (2015) and slightly adopted by Erkaev et al. (2020) on the abundance of NH$_3$ and the outgassing efficiency of N$_2$. They greyish and reddish area gives the parameter space within which the present-day atmosphere would be consistent with all constraints. Glein (2015) did not include atmospheric escape into his model, while Erkaev et al. (2020) did. This reduces the parameter space for an initially outgassed atmosphere.**

substantially true, then the study of Erkaev et al. (2020) further suggests that either

i) for a moderate to fast rotating young Sun the outgassing of Titan's atmosphere did not start before 20-67 million years to 425-495 million years after the formation of the solar system, respectively, or

ii) that the young Sun was a slow rotating young G-type star.

The reason for this is the following. Besides the outgassing efficiency ε, an important variable in the model of Glein (2015) is the so-called reaction progress variable $\xi_{N2}$, which describes the yield of N$_2$ from its precursor NH$_3$. If $\xi_{N2}$ = 1 then all endogenic NH$_3$ would have needed to be decomposed to N$_2$ via 2NH$_3 \rightarrow$ N$_2$ + 3H$_2$. Values higher than $\xi_{N2}$ = 1 are therefore clearly not possible (see Fig. 4). However, if Titan's atmosphere already started outgassing ~4.5 Gyr ago, the thermal escape produced by the strong EUV flux of a fast rotating Sun would have been so strong that $\xi_{N2}$ would have to be as high as up to $\xi_{N2}$ ~ 11. By considering the model of Glein (2015) and therefore assuming an endogenic origin of Titan's atmosphere, Erkaev et al. (2020) restricted the initial fractionation of $^{14}N/^{15}N$ to be <172, which is close to the present-day value and in agreement with an origin from ammonia ices (e.g. Mandt et al. 2014). The red, green, and blue shaded areas in Fig. 3b illustrate the earliest possible times for the outgassing of Titans nitrogen atmosphere for a slowly, moderately, and fast rotating young Sun according to Erkaev et al. (2020).

It further has to be noted that some recent studies found that the Sun originated most likely as a slow to moderate rotating star (Odert et al., 2018; Saxena et al., 2019; H. Lammer et al., 2020; e.g. Johnstone et al., 2020). Lammer et al. (2020) found that the present-day $^{36}Ar/^{38}Ar$ and $^{20}Ne/^{22}Ne$ ratios in Venus and Earth's atmosphere cannot be reproduced if the Sun originated as a moderate to fast rotator. In addition, another study by Johnstone et al. (2020) found that the terrestrial atmosphere would not have been stable during the Hadean and early Archean eons under the strong EUV flux of a fast rotating Sun, which is also in agreement with the work of Saxena et al. (2019), who found that the Sun likely has been a slow rotator based on sodium and potassium constraints from the Lunar regolith. These findings indicate that the nitrogen atmosphere of Titan could theoretically have been outgassed already very early after the formation of the solar system and not only after about 500 million years as it would have been the case for a fast rotating young Sun (Erkaev et al. 2020). A later outgassing, however, is not excluded by this study.

Another clue to the origin of Titan's atmosphere can be drawn from the atmospheric abundances of the noble gases (see e.g. Hörst, 2017). Since the ratio of $^{36}Ar$ to $^{14}N$ is orders of magnitudes below solar (Niemann et al., 2005, 2010), this is an indication that only an unsubstantial amount of $N_2$ could have been accreted during the formation of the Saturnian moon, since $^{36}Ar$ is trapped in ices at much colder temperatures than $N_2$ (Bar-Nun et al., 1988; Owen and Bar-Nun, 1995; Niemann et al., 2005). However, aerosols in Titan's atmosphere can trap $^{36}Ar$, which might remove enough 36Ar from its atmosphere over its lifetime to explain the present-day

[36]Ar/N$_2$ ratio (Jacovi and Bar-Nun, 2008). But the existence of the radiogenic [40]Ar in the atmosphere of Titan further suggests that outgassing from the interior played a major role in Titan's history (Niemann et al., 2005) which might be another indication for a later origin of its atmosphere through outgassing from the interior. An origin from the protosolar nebula seems to also be confirmed by the CO-chemistry in Titan's atmosphere (Hörst et al., 2008; Hörst, 2017 and references therein). Concerning its age, interior models further indicate that major outgassing mainly took place during the last ~1 Gyr (Tobie et al., 2012, Castillo-Rogez and Lunine, 2010), which is supported by the [12]C/[13]C ratio in CH$_4$ (Mandt et al., 2012; Nixon et al., 2012).

Determining the noble gas abundances and its isotopic ratios should give further clues on whether Titan's nitrogen originated from NH$_3$ in its interior or from another source. The endogenic model by Glein (2015) predicts quite specific ratios of [20]Ne/[22]Ne $\approx$ 8.9 and [36]Ar/[38]Ar $\approx$ 5.3 based on their outgassing efficiencies and on their potential origin. The latter was tentatively determined by the Cassini-Huygens/GCMS instrument (Niemann et al., 2010) to be $\approx$ 5, (Niemann et al., 2010), but this will have to be confirmed by future missions. In a more recent article, Glein (2017) also speculates on a nebula origin of Titan's Ne, which might be another explanation of its potentially close-to solar [22]Ne/[36]Ar ratio of about (0.05 – 2.5) times the Sun's value (Niemann et al., 2010), and would not be in contradiction with a non-solar origin of Titan's nitrogen (Glein, 2017). other noble gas ratio could so far, however, not be measured at Titan. Abundances or ratios of noble gases in comets, which could give another clue on the origin of Titan's atmosphere, were by now not measured at all.

Further isotopic measurements at Titan itself, can be performed by the upcoming Dragonfly mission, which arrives at the satellite in the mid-2030s to investigate its building blocks (Lorenz et al., 2018; Turtle et al., 2018). It consists of a lander that uses rotors for flying to different sites, and will carry on-board the Dragonfly Mass Spectrometer DraMS (Lorenz et al., 2018). This will perform Laser Desorption/ionization Mass Spectrometry (LDMS) and Gas Chromatography Mass Spectrometry (GCMS) and uses elements of Curiosity's SAM-suite (Mahaffy et al., 2012) and the Exomars/MOMA instrument (Goesmann et al., 2017). Since this mission will also study Titan's surface, it might be able to attain the $^{14}NH_3/^{15}NH_3$ ratio in its ices, a crucial measurement for determining the origin of Titan's atmospheric $N_2$, an issue that is still not entirely settled.

A recent study by Miller et al. (2019) proposes that accreted refractory organics may contribute up to 50% to Titan's present-day atmosphere. They argue that the $^{14}N/^{15}N$-ratio of cometary ammonia (~136; Shinnaka et al. (2016)) is lower than in Titan's atmosphere whereas cometary Ar/$N_2$-ratios seem to be too high (Miller et al., 2019). Refractory organics on the other hand show higher mean $^{14}N/^{15}N$-ratios (K. D. McKeegan et al., 2006) and lower mean Ar/$N_2$-ratios (Lodders, 2010) than Titan's atmosphere. Miller et al. (2019) therefore conclude that the isotopic and noble gas constraints of Titan suggest $NH_3$ together with refractory organics were the dominant sources for its atmospheric $N_2$. They further conclude that this mixture may have been the result of an accretion of material from the warm Saturnian sub-nebula followed by the heating of organics in Titan's interior and a subsequently outgassing into the atmosphere. It also has to be noted that photochemical fractionation, which tends to remove the heavier isotope from the atmosphere (Krasnopolsky, 2016), likely was not the reason for the fractionation from the cometary value to the present-day value according to Miller et al. (2019), since this process can only be relevant with methane being abundant in the atmosphere. But methane, however, might have been depleted in Titan's atmosphere early on, most likely as long as 1 Gyr ago or even longer (Yung et al., 1984; Mandt et al., 2012). Interior models (Tobie et al., 2006), on the other hand, also suggest earlier episodic outgassing of methane, so there could have been extended periods of methane in the atmosphere even before 1 Gyr ago which could have been a driver of photochemical fractionation.

A further insight into the origin of Titan's $N_2$ might be revealed by a measurement of the $^{14}N/^{15}N$-ratio of $N_2$ in Enceladus' plumes (Glein et al., 2009) which is expected to originate from the oxidation of $NH_3$ into $N_2$ (Matson et al., 2007; Glein et al., 2008). Since $N_2$ at Enceladus is not affected by fractionation through atmospheric escape this might also lead to a better understanding of the initial $^{14}N/^{15}N$-ratio of Titan and hence of its origin (Glein et al., 2009).

## 4 Pluto

That Pluto (Fig. 6Fig. 6) possesses an atmosphere was suggested as early as 1974 (Hart, 1974; Golitsyn, 1975; Cruikshank and Silvaggio, 1980; Fink et al., 1980; Trafton and Stern, 1983; Brosch, 1995) and confirmed via ground-based observations through stellar occultation's in 1988 (Hubbard et al., 1988; Elliot et al., 1989). These and further studies (e.g. Young et al., 1997; Lellouch et al., 2011; Olkin et al., 2014; Bosh et al., 2015; Dias-Oliveira et al., 2015) showed that the main constituent of its atmosphere was $N_2$ which was recently confirmed by the New Horizons space mission (Stern et al., 2015; Gladstone et al., 2016). Contrary to Titan, however, Pluto's tenuous atmosphere only has a surface pressure of about 6-24 μbar depending on its orbital location (Young et al., 1997; Lellouch et al., 2011). But as for Titan, a determination of Pluto's atmospheric $^{14}N/^{15}N$-ratio might reveal important hints on its origin (Jessup et al., 2013; Mandt et al., 2016). Observations with the Atacama Large Millimeter Array (ALMA) of $HC^{14}N$ and a non-detection of $HC^{15}N$ in Pluto's atmosphere provide a lower limit of 125 for $^{14}N/^{15}N$ in HCN (Lellouch et al., 2017). Although a pre-New Horizons study by Mandt et al. (2016) showed that if Pluto's nitrogen originated as $N_2$ its ratio could not be lower than 324 whereas it would be less than 154 if it would have been originated from ammonia. Updated work based on the observations made by New Horizons (Gladstone et al., 2016; Young et al., 2018) showed that the uncertainties in how condensation of HCN fractionates the isotopes creates large uncertainties for how to interpret a current ratio in $N_2$ (Mandt et al., 2017).

Due to the low gravity of Pluto it was expected that its atmospheric $N_2$ would easily escape. While it was debated whether this escape might be hydrodynamic (Krasnopolsky, 1999; Tian and Toon, 2005; Strobel, 2008) with loss rates of up to $10^{27} - 10^{28}$ molecules s$^{-1}$ (Tian and Toon, 2005; Zhu et al., 2014), or enhanced Jeans escape (Tucker et al., 2011), New Horizons showed that its atmosphere is currently lost via Jeans escape with the loss rate being in the range of ~$10^{23}$ molecules s$^{-1}$. Based on the pre-New Horizons escape models, Singer and Stern (2015) suggested that Pluto's $N_2$ should be of endogenic origin, if the estimated escape rates would have been so high that cometary impacts could not have resupplied its atmospheric $N_2$. In this case internal processes would be needed to outgas $N_2$ from the interior into the atmosphere (Singer and Stern, 2015). A post-New Horizons model by Glein and Waite (2018) considered whether the reservoir of Pluto's $N_2$ might be of primordial origin for which they developed two different model approaches, i.e. i) a "cometary model" which assumes a composition similar to comets with trapped $N_2$ from the solar nebula, and ii) a "solar model" which assumes a solar abundances of $N_2$. They found that, while their cometary model can account for the amount of $N_2$ stored in Sputnik Planitia, their solar model can provide such a large inventory of $N_2$ that strong atmospheric escape could have been possible in the past. Glein and Waite (2018) therefore conclude that the origin of Pluto's nitrogen inventory is likely primordial, i.e. accreted early as $N_2$ and might not originate from a secondary source. However, a solar as well as a cometary origin would result in a CO/$N_2$ ratio being significantly higher than the atmospheric ratio as measured by New Horizons and ALMA; Glein and Waite (2018) suggest a burial mechanism that depletes CO from Pluto's atmosphere to solve this paradox.

Even though New Horizons showed that the present-day escape rates are in the Jeans escape regime, it is important to determine whether the dominant escape process over Pluto's history was hydrodynamic or Jeans escape, as the studies of Singer and Stern (2015) and Glein and Waite (2018) illustrate. Since both processes fractionate very differently (e.g. Volkov et al., 2011; Mandt et al., 2014), measuring Pluto's atmospheric $^{14}$N/$^{15}$N-ratio would be crucial to not only gain better insights into its origin but also into its escape mechanisms. In addition, constraining the ratio of $^{36}$Ar/N$_2$ at Pluto's surface can give an important constraint on whether its N$_2$ really is primordial. A high amount of $^{36}$Ar would be consistent with the primordial

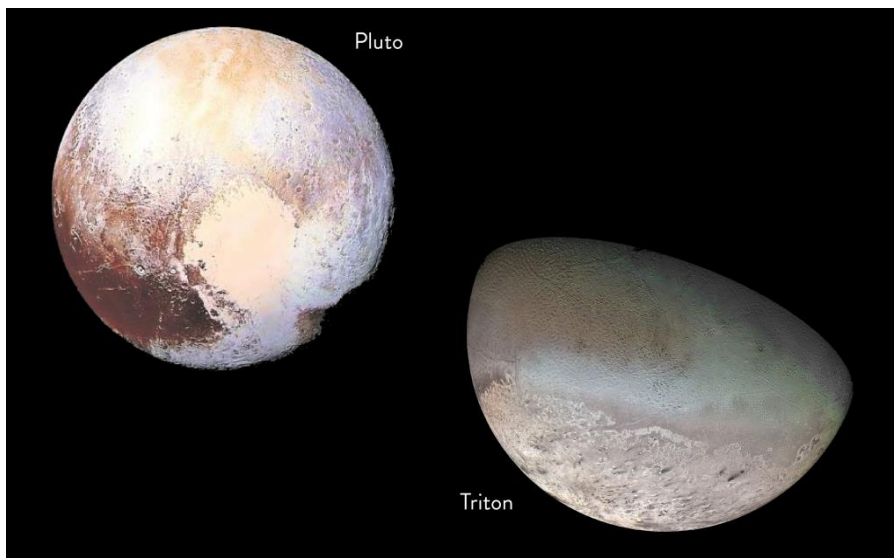

**Fig. 6. Left: New Horizons image of Pluto (Courtesy NASA/JHUAPL/SwRI); right: Voyager 2 image of Neptune's moon Triton (Courtesy NASA/Jet Propulsion Lab/U.S. Geological Survey).**

hypothesis, while low amounts would imply an origin from a secondary source (Glein and Waite 2018).

Further interesting questions in view of the evolution of Pluto's nitrogen atmosphere are how its escape and pressure vary over its elongated orbit. While the escape rate seems to change only by a factor of 2 (Tian and Toon, 2005), its atmosphere might even collapse due to cooling on the outer parts of its orbit (e.g. Hansen et al. (2015)). From 1988 to 2002, however, its atmospheric pressure increased by a factor of about two from $2.33 \pm 0.24$ µbar at $r = 1215$ km in 1988 (Yelle and Elliot, 1997) to $5.0 \pm 0.6$ µbar in 2002 (Sicardy et al., 2003) even though Pluto was receding from the Sun (Elliot et al., 2003; Sicardy et al., 2003). This effect might be due to Pluto's volatile cycle which is not only determined by its distance from the Sun but also by its high obliquity and therefore its subsolar latitude as was predicted by Hansen and Paige (1996). Meza et al. (2019), who measured the surface pressure of Pluto through eleven stellar occultation events between 2002 and 2016, further reported a continuous increase in pressure of up to $12.04 \pm 0.41$ µbar at the surface in 2015 (compared to $4.28 \pm 0.44$ µbar in 1988) and $6.92 \pm 0.076$ µbar at $r = 1215$ km. This rise is due to i) the heating of nitrogen ice in Sputnik Planitia and in the northern mid latitudes because of the direct insolation from the Sun, and ii) the orbital position of Pluto close to the Sun (Bertrand and Forget, 2016).

## 5 Triton

Neptune's moon Triton (Fig. 6) also has a tenuous nitrogen-dominated atmosphere similar to Pluto's. While the Voyager flyby of 1989 measured an $N_2$-surface pressure of about 14 µbar (Gurrola, 1995), it has risen to about 40 µbar in 2009 (Lellouch et al., 2010). First models of the present-day thermal nitrogen escape rates at Triton estimated values in the range of $10^{24} – 10^{25}$ s$^{-1}$ (Yung and Lyons, 1990; Summers and Strobel, 1991; Lyons et al., 1992; Krasnopolsky et al., 1993; Krasnopolsky and Cruikshank, 1995). Lammer (1995) found that non-thermal escape rates via sputtering through particles within the magnetosphere of Neptune could also lead to comparable values in the order of $5 \times 10^{24}$ s$^{-1}$. These escape rates are about one to two orders higher than the thermal escape rates that New Horizons found at Pluto. If, for thermal escape, one considers the closer location to the Sun during the Voyager 2 flyby, on which most of the Triton models are based, then such higher escape rates might indeed be likely. Also the value for sputtering at Triton seems reasonable if one compares it with Titan, for which the sputtering escape rate is estimated to be around $3 - 4 \times 10^{25}$ s⁻¹ (Shematovich et al., 2003; Michael et al., 2005). Within a more recent study, Strobel and Zhu (2017), however, modelled thermal escape rates from Triton's atmosphere that would be significantly higher, i.e. about 2.6 x $10^{27}$ molecules s⁻¹ which would also be orders of magnitudes higher than the escape from present-day Pluto's atmosphere. Whether this is likely or not could be resolved by a new mission to the Ice Giants.

As for Pluto, also in the case of Triton the atmospheric $^{14}$N/$^{15}$N-ratio is yet unknown and its origin unclear, but a similar source as Pluto's nitrogen-dominated atmosphere might be likely. Glein and Waite (2018) speculate that if Triton formed from the same reservoir than Pluto, then the non-detection of $CO_2$ at Pluto but its detection at Triton might be an indicator on some geochemical endowment or geophysical processes at Triton. The "solar model" by Glein and Waite (2018) interestingly might also provide sufficient $N_2$ for the idea of a massive early $N_2$ atmosphere around Triton that could have been accreted from the protosolar or proto-Neptunian nebula, as suggested by Lunine and Nolan (1992). They proposed that the lifetime of such a thick atmosphere could have been more than a billion years, depending on the escape rates. However, the current knowledge on the origin and evolution of Triton's atmosphere is poorly known and a new mission to Neptune or particularly to Triton, such as the recently proposed NASA flyby-mission Trident (Mitchell et al., 2019; Prockter et al., 2019), which carries a plasma spectrometer that will probe Triton's tenuous atmosphere by passing through its upper layers at about 500 km (Mitchell et al., 2019), would lead to new insights not only into the origin and evolution of Triton's atmosphere but also into Pluto's.

# 6 Conclusion

Within the last 15 years our knowledge of the nitrogen atmospheres of the icy bodies in the solar system significantly advanced particularly due to Cassini around Saturn and the flyby of New Horizons at Pluto. However, other space missions such as Rosetta and ground-based observations of the icy bodies and particularly of comets also helped to trace the origin of their atmospheric $N_2$. The different isotopic signatures of nitrogen in cometary ammonia ices, refractory organics, chondrites, and in the proto-solar nebula define different reservoirs of which the icy bodies could have accreted their nitrogen. In contrast to Earth – its nitrogen atmosphere most probably originated from the chondritic reservoir – it currently seems likely that their building blocks might have originated from $NH_3$ in comets and potentially from refractory organics in case of Titan, and from comets and/or the protosolar nebula in case of Pluto and potentially Triton. For understanding their origin, however, it is crucial to study the various fractionation processes that take place in the atmospheres of these bodies as several studies successfully did in the recent past. Here, it is important to note that the evolution of Titan, Triton and Pluto is strongly dependent on the evolution of the Sun which has profound effects on their atmospheres, promoting escape and chemical reactions, and therefore on isotopic fractionation. Only if we understand the solar evolution, will it be possible to trace back the origin of the nitrogen atmospheres of the icy bodies, and the history of the Sun can best be reconstructed via comparative planetology.

Table 3. Key measurements for a better understanding of the nitrogen atmospheres of icy satellites.

| key measurement | solar system body | scientific question |
| --- | --- | --- |
| $^{20}Ne/^{22}Ne$, $^{36}Ar/^{38}Ar$, $^{36}Ar/^{38}Ar$ and its abundances; Kr and Xe abundances | Titan, Pluto, Triton, comets | testing the endogenic model of Glein (2015) for Titan; clues on the building blocks of the atmospheres of Titan, Triton, and Pluto; origin of the noble gases on these bodies; history of atmospheric escape due to the fractionation of the lighter from the heavier isotope |
| $^{14}N/^{15}N$ in $N_2$ | Pluto, Triton | testing the origin of Pluto's and Titan's N2 (protosolar vs. cometary origin); clues on the history of atmospheric escape at Triton and Pluto |

| $^{14}N/^{15}N$ in $NH_3$ | Titan's surface ices | confirming the origin of Titan's $N_2$ from ammonia |
| $^{14}N/^{15}N$ | Saturn | determining the isotopic ratio of Saturn and its subnebula's as clues to Titan's origin |
| $^{14}N/^{15}N$ | Enceladus' plumes | comparing Enceladus hydrothermal plumes with Titan's composition |
| radiogenic $^{129}Xe$ and $^4He$ | Titan | constraining the outgassing history of Titan |
| atmospheric escape of nitrogen | Triton | retrieving escape rates for Triton's atmosphere for better understanding its stability |

Even though comparative planetology is helpful to understand the origin and evolution of nitrogen atmospheres in general, future in-situ missions to the icy bodies will be required to answer currently unknown questions and to better understand their origin (see Table 3 for future key measurements for a better understanding of the nitrogen atmospheres of the icy satellites in the solar system, and, particularly, Glein (2015) for a thorough analysis on future measurements for a better understanding of Titan). In 2019 NASA announced that the space mission Dragonfly (Lorenz et al., 2018; Turtle et al., 2018) which is going to be funded for flying to Titan. The mission will start in 2026 and arrive in 2034. Its main goal is to study Titan's origin and habitability. It will explore the building blocks of the moon in unprecedented detail and will also tell us a lot about the origin of the other icy bodies in the solar system such as Triton and Pluto. Our knowledge about Triton, however, is mainly based on the flyby of Voyager 2 and on comparative planetology. Here, a mission to Neptune and its moon such as the recently proposed NASA Trident mission (Prockter et al., 2019) – or even a twin ice giants mission (see e.g. Hofstadter et al. (2017), ESA's CDF Study Report on Ice Giants (2019) and Fletcher et al. (2019) for the scientific potential of such missions) - would be crucial to significantly enhance our scientific knowledge about this satellite, its tenuous nitrogen atmosphere and about icy bodies in general. Finally, another mission that was recently selected by ESA, will help to better understand the potential building blocks of Titan, Pluto, and Triton. Comet Interceptor (Snodgrass and Jones, 2019), which will launch in 2028, will be the first spacecraft to visit a pristine comet.

**Acknowledgments**   MS acknowledges the support of Europlanet 2020 RI. Europlanet 2020 RI has received funding from the European Union's Horizon 2020 research and innovation programme under grant agreement No 654208. NVE and HL acknowledge the FWF NFN Project S11607-N27. NVE acknowledges RFBR grant No 18-05-00195-a. KM acknowledges NASA grants 80NSSC18K1233 and 80NSSC19K1306. BM acknowledges the European Research Council grant 695618. We finally thank an anonymous referee who helped to significantly enhance the value of our review.